\documentclass[journal]{IEEEtran}
\usepackage[colorlinks]{hyperref}

\usepackage{cite}
\usepackage{amsthm}
\usepackage[cmex10]{amsmath}
\usepackage{mathtools}
\usepackage{array}
\usepackage{bm}

\usepackage{multicol,booktabs,tabularx}

\usepackage{cases}
\usepackage{algorithm}
\usepackage[tight,footnotesize]{subfigure}
\usepackage{amssymb}
\usepackage{array}
\usepackage{multirow}

\begin{document}

\title{\LARGE Context-Aware Deep Learning for Robust Channel Extrapolation in Fluid Antenna Systems}

\author{Yanliang Jin, 
            Runze Yu, 
            Yuan Gao, 
            Shengli Liu, 
            Xiaoli Chu,~\IEEEmembership{Senior Member,~IEEE}, \\
            Kai-Kit Wong,~\IEEEmembership{Fellow,~IEEE}, and 
            Chan-Byoung Chae, \emph{Fellow, IEEE}
\vspace{-5mm}

\thanks{The work of Y. Jin, R. Yu, Y. Gao and S. Liu is supported by Shanghai Natural Science Foundation under Grant 25ZR1402148.} 
\thanks{The work of K. K. Wong is supported by the Engineering and Physical Sciences Research Council (EPSRC) under Grant EP/W026813/1.}
\thanks{The work of C.-B. Chae was in part supported by the Institute for Information and Communication Technology Planning and Evaluation (IITP)/NRF grant funded by the Ministry of Science and ICT (MSIT), South Korea, under Grant RS-2024-00428780 and 2022R1A5A1027646.}

\thanks{Y. Jin, R. Yu, Y. Gao and S. Liu are with the School of Communication and Information Engineering, Shanghai University, China (e-mail: $\rm \{jinyanliang,urleaves,gaoyuansie,victoryliu\}@shu.edu.cn$).}
\thanks{X. Chu is with the Department of Electronic and Electrical Engineering, the University of Sheffield, UK (e-mail: $\rm x.chu@sheffield.ac.uk$).}
\thanks{K. K. Wong is with the Department of Electronic and Electrical Engineering, University College London, WC1E 7JE, United Kingdom, and also with Yonsei Frontier Lab, Yonsei University, Seoul, South Korea (e-mail: $\rm kai\text{-}kit.wong@ucl.ac.uk$).}
\thanks{C.-B. Chae is with the School of Integrated Technology, Yonsei University, Seoul, 03722 South Korea (e-mail: $\rm cbchae@yonsei.ac.kr$).}

\thanks{Corresponding author: Yuan Gao.}
}

\maketitle

\begin{abstract}
Fluid antenna systems (FAS) offer remarkable spatial flexibility but face significant challenges in acquiring high-resolution channel state information (CSI), leading to considerable overhead. To address this issue, leveraging the inherent spatial continuity of electromagnetic fields, we propose CANet, a robust deep learning model for channel extrapolation in FAS. CANet combines context-adaptive modeling with a cross-scale attention mechanism and is built on a ConvNeXt v2 backbone to improve extrapolation accuracy for unobserved antenna ports. \textcolor{red}{To further enhance robustness, we introduce a novel spatial amplitude perturbation strategy, inspired by frequency-domain augmentation techniques in image processing. This motivates the incorporation of a Fourier-domain loss function, capturing frequency-domain consistency, alongside a spectral structure consistency loss that reinforces learning stability under perturbations.} Our simulation results demonstrate that CANet not only outperforms benchmark models in terms of estimation accuracy across a wide range of signal-to-noise ratio (SNR) levels and carrier frequencies, but also significantly reduces the outage probability with moderate computational complexity.
\end{abstract}


\IEEEpeerreviewmaketitle

\section{Introduction}
\IEEEPARstart{W}{ith} the advent of sixth-generation (6G) mobile communications, the demand for higher capacity, wider coverage, improved positioning and greater flexibility in wireless systems is accelerating\cite{xu2025enhanced,gao2025stochastic,gao2024performance,du2024secure}. Meeting these ambitious requirements necessitates the development of advanced physical-layer technologies. Among them, the fluid antenna system (FAS) \cite{Wong2022Bruce,New2024Tutorial,SSnet2025gao} has emerged as a promising antenna architecture that introduces new spatial degrees of freedom (DoF) into wireless communication. FAS represents the capability of shape-and-position-reconfigurable antenna technology to empower the physical layer. By leveraging mechanical motion \cite{I27_basbug2017design}, liquid metals \cite{I24_shen2024design}, reconfigurable pixel-based structures \cite{I26_zhang2024pixel} or metamaterials \cite{Liu-2025arxiv,Lu-2025}, FAS enables dynamic port selection from a massive number of predefined locations. This adaptability facilitates spatial diversity gains in nearly continuous space and enhances the system's capability to respond intelligently to the  wireless channel. Since the introduction of FAS by Wong {\em et al.}~in \cite{I22_wong2020perflim,I20_wong2021FAS}, there have been lots of efforts in understanding the achievable performance of FAS \cite{G5_new2023SISO-FAS,H7_Espinosa2024Anew}, showing potential for reconfigurable intelligent surface (RIS) systems \cite{G16_Lai2024FAS-RIS}, index modulation \cite{G27_Zhu2024FA-IM,G26_Zhu2024FAIM-RIS}, wideband communication \cite{H11_hong2025Downlink}, and integrated sensing and communication (ISAC) \cite{Wang2024Fluid}. 

Nonetheless, the promising performance of FAS relies on the availability of CSI \cite{gao2026csiextra,G11_ye2024MIMO-FAS,G13_Efrem2024MIMO-FAS,New2024Information}, and the high-resolution nature of FAS also leads to substantial overhead in acquiring CSI, including pilot overhead, feedback latency, and energy consumption\cite{jiang2025towards,OTFSGao2025}. In recent years, several studies explored the integration of physical modeling and data-driven approaches for CSI prediction \cite{gao2025enabling,gao2024c2s,jin2025linformer}. For instance, \cite{Li2024Model} proposed an inference algorithm to enable high-accuracy extrapolation in a massive FAS using only a small number of observed ports. Additionally, in \cite{Wu2024Channel}, an extrapolation scheme based on masked language models, which reconstructs the full channel from a limited number of observed port CSI, was proposed. Going one step further, the approach in \cite{zhang2024learning} integrates the diffusion mechanism of graph neural networks with an asymmetric masked autoencoder architecture, significantly improving both extrapolation accuracy and flexibility in high-resolution FAS settings. Prior to these, there were attempts that exploited channel sparsity \cite{New-2025twc} and used Bayesian approaches \cite{Xu-2025wcl}. CSI prediction for multiuser FAS using deep learning was also reported in \cite{Waqar2023Deep}. 

\textcolor{red}{However, the existing research shows limitations in reliable CSI acquisition for FAS. Due to the extremely dense deployment of candidate ports, full-port CSI acquisition via sequential switching incurs prohibitive latency and pilot overhead, making extrapolation from a highly limited number of observed ports inevitable. The core challenge lies in reconstructing high-resolution spatial correlation from sparse and noise-contaminated observations, under which conventional learning-based models are prone to noise overfitting and poor generalization to unobserved ports.}

\textcolor{red}{Motivated by these challenges, we propose CANet, a deep learning framework specifically designed for robust channel extrapolation in FAS. CANet adopts a synergistic local-to-global architecture with multi-domain regularization, integrating a context-adaptive block (CAB) for reliable local feature fusion, a cross-scale contextual attention (CSCA) mechanism for global feature completion, and a ConvNeXt v2 backbone. In addition, a spatial amplitude perturbation strategy and \textbf{a Fast Fourier Transform (FFT)-based} loss are introduced to enhance robustness against noise by enforcing frequency-domain consistency.}

\section{Channel Extrapolation for FAS}
\textcolor{red}{Consider a slow fluid antenna multiple access (FAMA) downlink system\cite{Waqar2023Deep}} where a base station (BS) with $M_t$ fixed-position antennas (FPAs) communicates with $U_t$ user terminals on the same time-frequency resource unit. Each user terminal employs a 2D FAS with $N_s=N_{x}\times N_y$ switchable positions (or ports), which are evenly distributed over a planar surface of physical dimensions $W_s=W_{x}\times W_y$, \textcolor{red}{where each user activates only one port in each coherence block.} In a 3D rich scattering environment, the spatial correlation between port $(n_x^s,n_y^s)$ and port $(\tilde n_x^s,\tilde n_y^s)$ is given by
\begin{small}\begin{equation}\label{eq:Jelement}
\begin{array}{l}
J_{\left(n_{x}, n_{y}\right),\left(\tilde{n}_{x}, \tilde{n}_{y}\right)} \\
\quad=j_{0}\left(\frac{2 \pi}{\lambda} \sqrt{\left(\frac{\left|n_{x}-\tilde{n}_{x}\right|}{N_{x}-1} W_{x}\right)^{2}+\left(\frac{\left|n_{y}-\tilde{n}_{y}\right|}{N_{y}-1} W_{y}\right)^{2}}\right),
\end{array}
\end{equation}\end{small}where $j_0(\cdot)$ denotes the spherical Bessel function of the first kind, which can be approximated by $\mathrm{sinc}(x) = {\sin x}/{x}$, and $\lambda$ represents the carrier wavelength. The same spatial correlation matrix $\mathbf{J}_s = \mathbf{U}_s \mathbf{\Lambda}_s \mathbf{U}_s^H$ is shared by all users, where $\mathbf{U}_s\in \mathbb{C}^{N_s \times N_s}$ and $\mathbf{\Lambda}_s\in \mathbb{C}^{N_s\times N_s}$ denote the eigenvector matrix and the diagonal eigenvalue matrix, respectively.

For a given user, the downlink channel over all FAS ports and all FPAs can be modeled as
\begin{small}\begin{equation}\label{eq:channel_typical}
    \mathbf{g} = \delta\, \mathbf{U}_s\sqrt{\mathbf{\Lambda}_s}\mathbf{G},
\end{equation}\end{small}
where $\mathbf{g}\in \mathbb{C}^{N_s\times M_t}$ denotes the small-scale fading channel matrix, $\mathbf{G}\in \mathbb{C}^{N_s\times M_t}$ consists of independent complex Gaussian random variables with zero mean and variance $1/2$ per real dimension, and $\delta$ characterizes the large-scale path loss.

For the $u$-th user terminal, $u\in\{1,\dots,U_t\}$, its channel realization over all FAS ports and BS antennas is denoted by $\mathbf{g}_u\in\mathbb{C}^{N_s\times M_t}$, which follows the same statistical model as in \eqref{eq:channel_typical}. \textcolor{red}{As demonstrated in \cite{Alr2019Deep}, if the position-to-channel mapping is bijective in a static FAS environment (with fixed geometry, port locations, and scattering conditions), then a channel-to-channel mapping also exists. In other words, once the environment is fixed, each FAS port position almost surely corresponds to a unique downlink channel vector, and vice versa. Unleashing the full potential of FAS necessitates the acquisition of $\mathbf{g}_u$. Conventional pilot-based approaches to acquire $\mathbf{g}_u$ cause excessive overhead \cite{zhang2024learning}. To this end, we propose channel extrapolation to acquire the CSI of a large set of ports $\widehat{\mathbf{g}}_{u,\mathcal{A}}$ using a small set of observable ports $\mathbf{g}_{u,\mathcal{B}}$, which is expressed as:}

\begin{small}\begin{equation}
    \widehat{\mathbf{g}}_{u,\mathcal{A}} = f(\mathbf{g}_{u,\mathcal{B}}),
\end{equation}\end{small}where $f(*)$ is the channel extrapolation function. The aim of the extrapolation is to minimize the expected normalized mean-square-error (NMSE) between the estimated CSI $\widehat{\mathbf{g}}_{u,\mathcal{A}}$ and the ground truth CSI ${\mathbf{g}}_{u,\mathcal{A}}$, which is expressed as:

\begin{small}\begin{equation}
     \min_{f(*)}\mathbb{E}\left( \frac{|\widehat{\mathbf{g}}_{u,\mathcal{A}}-{\mathbf{g}}_{u,\mathcal{A}}|^2}{|{\mathbf{g}}_{u,\mathcal{A}}|^2} \right).
\end{equation}\end{small}

\section{Proposed Method}
Here, we propose CANet, a noise-robust architecture for the channel extrapolation task. We design a CAB to locally fuse the features of known CSI with weighted integration, while constraining the propagation of features from unknown CSI. To enhance global modeling capability, we introduce a CSCA module to facilitate feature completion for large unobserved port regions. Moreover, we adopt ConvNeXt v2 with dropout regularization to improve feature extraction and generalization after feature completion \cite{Woo2023Convnext}. A spatial amplitude perturbation module is also incorporated to enhance the model's adaptability and robustness under various noise conditions.

\subsection{Masking Strategy}
We denote the ground truth CSI as \( \mathbf{g} \in \mathbb{C}^{N_s \times M_t} \), and construct a tensor \( \mathbf{u} \in \mathbb{R}^{2M_t \times N_y \times N_x} \) to represent the concatenated real and imaginary components of the CSI across a 2D fluid antenna grid of size \( N_x \times N_y \). To simulate partial observations, a subset of antenna ports is randomly masked along the spatial grid, with a masking ratio typically ranging from $80\%$ to $95\%$. The resulting masked tensor is denoted as \( \tilde{\mathbf{u}} \in \mathbb{R}^{2M_t \times N_y \times N_x} \), in which the unobserved positions are replaced with a distinct negative value (e.g., $-10$) to make them distinguishable for the network. Additionally, we define a binary indicator matrix \( \mathbf{E}_{\text{flag}} \in \{0,1\}^{N_y \times N_x} \), where a value of $1$ indicates an observed port and $0$ is a masked (unobserved) port. No prior knowledge of the channel is used to select ports for extrapolation. Instead, a random masking strategy is employed to ensure flexibility and enhance the generalization capability of the model.

\begin{figure*}[]
\centering
\includegraphics[width=0.75\linewidth]{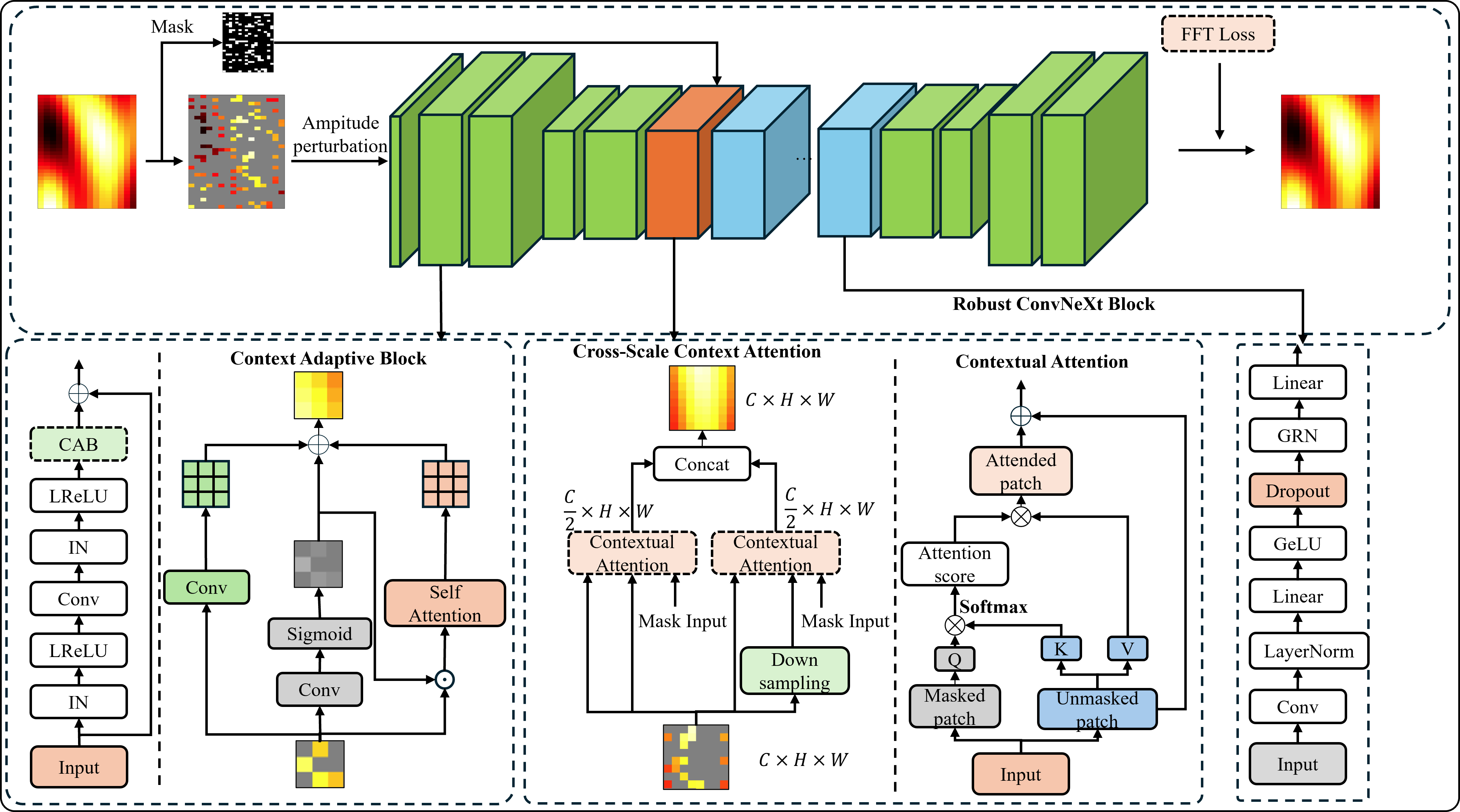}
\caption{The proposed CANet for CSI extrapolation in FAS channels, where $\oplus$, $\otimes$ and $\odot$ denote fuse, dot product and element-wise product, respectively.}\label{fig:tsppo}
\vspace{-2mm}
\end{figure*}

\subsection{Model Architecture}
\textcolor{red}{The architecture of CANet is designed to exploit the spatial continuity and spectral redundancy of FAS. It leverages a local-to-global inductive bias to reconstruct CSI from sparse observations, as detailed below.}
\subsubsection{CAB}
\textcolor{red}{To tackle the challenge of irregular sparse observations inherent in FAS, we design the CAB to precisely extract effective information within the receptive field. Standard convolution treats all pixels equally, making it difficult to distinguish valid signals from the values at unobserved positions. To overcome this, CAB employs a context-aware mechanism that dynamically identifies and weighs valid features from observed ports, ensuring accurate local feature extraction.}

Specifically, we first compute a soft mask (confidence map) over the entire feature map using convolution. This soft mask reflects the reliability of each feature.\footnote{Features affected by masked ports are assigned lower confidence, while those from observed regions receive higher confidence.} Then we perform a weighted fusion of the outputs from the convolutional and attention branches to further refine feature extraction within the receptive field. For each pixel $x_{o}\in\mathbb{R}^{C}$ on the feature map $F\in\mathbb{R}^{C\times H\times W}$, its corresponding soft mask value $m$ is computed by the following convolutional operation:
\begin{small}\begin{equation}
m = \sigma\left( \sum_{i\in \mathcal{B}_{o} }w _{i}^{m}x_{i}\right),
\end{equation}\end{small}
where $\sigma$ denotes the sigmoid activation function, $x_{i}$ represents the input features within the receptive field, $w_{i}$ denotes learnable convolutional filters, and $\mathcal{B}_{o}$ represents the receptive field range corresponding to filter $w^{m}$. The magnitude of $m$ signifies the confidence level of the feature at that location. Conditioned on the learned soft mask $m$, we then introduce a self-attention mechanism \cite{jin2025linformer} to derive a more contextually relevant feature representation $x^{a}$, given by
\begin{small}\begin{equation}
x^{a}
= \sum_{i\in \mathcal{B}_{o}}^{}\left\langle mw^{q}x_{o},mw^{k}x_{i} \right \rangle mx_{i}w^{v}x_{i},
\end{equation}\end{small}where $\left\langle\cdot,\cdot\right \rangle $ denotes the composite operation of dot-product with softmax. This self-attention mechanism empowers the model to focus on context information pertinent to the current pixel, thereby enhancing the capability of features to capture and discriminate CSI. In parallel, within the convolutional branch, we employ another set of convolution kernels $w^{c}$ to extract local channel features $x^{c}$ by
\begin{small}\begin{equation}
x ^{c} = \sum_{i\in \mathcal{B}_{o} }w _{i}^{c}x_{i}.
\end{equation}\end{small}

Finally, we perform a weighted fusion of the convolutional branch output $x_c$ and the attention branch output $x_a$ based on the soft mask $m$, yielding the output
\begin{small}\begin{equation}
o = m\cdot  x ^{c} + (1-m)\cdot x ^{a}.
\end{equation}\end{small}

\subsubsection{CSCA}
\textcolor{red}{Considering the finite aperture of FAS, multipath propagation creates specific geometric constraints and long-range correlations across the antenna surface. We introduce CSCA to capture these global dependencies that local convolutions might miss. By facilitating cross-scale contextual attention, it promotes global consistency of the extrapolated channel with the underlying propagation geometry over the finite aperture.} We first reduce the input feature channels to half of the original before feeding them into two self-attention modules. One branch directly undergoes self-attention at the same scale, while the other duplicates the feature map and down-samples it to half of the original resolution. The downsampled feature map is then processed along with the original-scale features within the self-attention mechanism. At each self-attention stage, patches are extracted from the input features, and the cosine similarity between patches in the missing region and those in the external region is computed to establish cross-region feature correlations. {Formally, let $q_i$ denote the $i$-th patch from the masked region $\psi^l$, and $k_j$ denote the $j$-th patch from the observed region $\phi^l$. The cosine similarity is computed as}
\begin{small}\begin{equation}
s_{i,j} = \left\langle \frac{q_i}{\|q_i\|}, \frac{k_j}{\|k_j\|} \right\rangle.
\end{equation}\end{small}
The attention weights are then computed by
\begin{small}\begin{equation}
a_{i,j}=\frac{\exp(s_{i,j})}{\sum_{j} \exp(s_{i,j})}.
\end{equation}\end{small}
The unobserved port patches are then reconstructed based on the computed attention scores, given by
\begin{small}\begin{equation}
\hat{q}=\sum_{j}a_{i,j}v_j,
\end{equation}\end{small}
where $v_{j}$ represents the $j$-th patch extracted from features outside unobserved regions. Finally we concatenate the results of the two branches as the final result.

\subsubsection{ConvNeXt v2 with dropout regularization}
\textcolor{red}{After the initial completion by CAB and CSCA, we employ the ConvNeXt v2 \cite{Woo2023Convnext} backbone to perform deep feature refinement and learn robust high-level channel structures.} ConvNeXt is known for its strong representation capabilities, and ConvNeXt v2 further enhances this by introducing the global response normalization (GRN) module, which enables more expressive feature learning. In our design, we employ the ConvNeXt v2 block integrated with dropout regularization to prevent overfitting and ensure generalization under diverse noise conditions.
By randomly dropping a subset of neurons during training, dropout alters the effective network structure in each training epoch, leading to a degree of statistical independence across different training instances, thereby reducing prediction variance and enhancing generalization under diverse noise conditions.

\subsubsection{Spatial amplitude perturbation}
\textcolor{red}{Since FAS channel extrapolation is highly sensitive to measurement noise and multipath variations, we introduce a Spatial Amplitude Perturbation module to enforce robustness by simulating spectral uncertainty.} To enhance the model's robustness, we propose a spatial amplitude perturbation strategy at the input stage, inspired by frequency perturbation techniques in the image domain \cite{Liu2024Universal,Yang2025Single}. Specifically, the input CSI features are first zero-mean normalized to remove the direct current (DC) component, then transformed into the frequency domain via FFT. The amplitude spectrum is then perturbed in a stochastic manner based on a predefined probability, effectively simulating potential frequency interference and noise. This approach improves the model's adaptability to unseen distributions:
\begin{small}\begin{equation}\label{eq:patially_zero-mean_feature}
\tilde{\mathbf{z}} = \tilde{\mathbf{u}} - \frac{1}{HW} \sum_{h=1}^{H} \sum_{w=1}^{W} \tilde{\mathbf{u}}_{h,w}.
\end{equation}\end{small}
\begin{small}\begin{equation}
\hat{\mathbf{z}} = \mathcal{F}^{-1} \left( \mathcal{F}(\tilde{\mathbf{z}}) \cdot \left(1 + \boldsymbol{\gamma} \cdot \boldsymbol{\xi} \cdot \boldsymbol{\delta} \right) \right) + \bar{\mathbf{z}},
\end{equation}\end{small}
where \( \tilde{\mathbf{z}} \) is the spatially zero-mean feature, \( \hat{\mathbf{z}} \) is the final perturbed feature after reconstruction, {\( \bar{\mathbf{z}} \) denotes the spatial mean component removed in \eqref{eq:patially_zero-mean_feature}}, \( \mathcal{F}(\cdot) \) and \( \mathcal{F}^{-1}(\cdot) \) denote the 2D discrete Fourier transform and its inverse. The perturbation mask \( \boldsymbol{\delta} \in \{0,1\}^{H \times W} \) above is sampled from a Bernoulli distribution with probability \( \mu \), and the amplitude coefficients \( \boldsymbol{\xi} \in [0,1]^{H \times W} \) are sampled from a normal distribution \( \mathcal{N}(0, 1) \). The hyperparameters \( \mu \) and \( \gamma \) control the perturbation probability and maximum strength, respectively. Only the amplitude of each frequency component is perturbed, while the phase is fixed to maintain structural consistency.

\subsubsection{Loss Function}
Since the CSI tensor is represented by concatenating the real and imaginary parts, we reconstruct it into a complex-valued tensor before computing the frequency-domain loss to enable the 2D Fourier transform. We denote the original CSI tensor as \( \mathbf{{U}_c} \in \mathbb{R}^{M_t \times N_y \times N_x} \), and \( \hat{\mathbf{{U}_c}} \in \mathbb{R}^{M_t \times N_y \times N_x} \) denote the predicted CSI recovered from the masked input $\tilde{\mathbf{{U}_c}}$. Let \( \Omega \subset \{1, \dots, N_y\} \times \{1, \dots, N_x\} \) denote the set of unobserved spatial locations, with \( |\Omega| = N_a \). The total loss is then defined as
\begin{small}\begin{equation}
\mathcal{L}_{\text{total}} = \mathcal{L}_{\text{MSE}} + \beta \mathcal{L}_{\mathrm{FFT}},
\end{equation}\end{small}
where $\mathcal{L}_{\text{MSE}}$ and $\mathcal{L}_{\mathrm{FFT}}$ \textbf{are the mean squared error (MSE) loss} and the FFT loss attributed to spatial amplitude perturbation, respectively, calculated as
\begin{small}\begin{align}
\mathcal{L}_{\text{MSE}} &= \mathbb{E} \left[ \frac{1}{|\Omega|} \sum_{(y,x) \in \Omega} \left\| \hat{\mathbf{{U}_c}}[:, y, x] - \mathbf{{U}_c}[:, y, x] \right\|_2^2 \right],\\
\mathcal{L}_{\mathrm{FFT}}& = \mathbb{E} \left[ \left\| \left| \mathcal{F}(\hat{\mathbf{{U}_c}}) \right| - \left| \mathcal{F}(\mathbf{{U}_c}) \right| \right\|_2^2 \right],
\end{align}\end{small}
in which \( \beta \) balances the contribution of the two losses. The FFT loss enforces global consistency in the spatial frequency domain, which helps preserve large-scale channel structure.

\vspace{-2mm}
\section{Numerical Results}
\subsection{Simulation settings}
\begin{table}[t]
\centering
\caption{Architecture of the proposed CANet.}
\label{tab:network_layers}
\small
\setlength{\tabcolsep}{3pt}
\renewcommand{\arraystretch}{0.95}
\begin{tabular}{l l c c c c}
\toprule
\textbf{Layers} & \textbf{Operator} & \textbf{Out} & \textbf{Ch.} & \textbf{K(S)} & \textbf{Act.} \\
\midrule
1      & CAB                 & $32{\times}16$ & 64  & $5(1)$ & LReLU \\
2      & CAB                 & $32{\times}16$ & 128 & $3(2)$ & LReLU \\
3      & CAB                 & $16{\times}8$  & 128 & $3(1)$ & LReLU \\
4--5   & CAB $\times 2$       & $16{\times}8$  & 256 & $3(1)$ & LReLU \\
6      & CSCA                & $16{\times}8$  & 256 & --     & --    \\
7--16  & ConvNeXt v2 $\times10$ & $16{\times}8$  & 256 & $7(3)$ & GELU  \\
17     & BilinearUp          & $32{\times}16$ & 256 & --     & --    \\
18--19 & CAB $\times 2$       & $32{\times}16$ & 128 & $3(1)$ & LReLU \\
20--21 & CAB $\times 2$       & $32{\times}16$ & 64  & $3(1)$ & LReLU \\
22     & Conv                & $32{\times}16$ & 16  & $7(1)$ & --    \\
\bottomrule
\end{tabular}

\vspace{0.5mm}
\footnotesize\emph{Note:} K: kernel size; S: stride; LReLU: LeakyReLU; GELU: Gaussian Error Linear Unit.
\end{table}

In the simulations, the FAS channel involves \( U_t \) users. Each user has a planar FAS of size \( W_s\,~\text{cm}^2 \) and \( N_s \) antenna ports. We assume \( M_t = U_t \), and the BS is equipped with \( M_t \) FPAs. The carrier frequency is $3.4~{\rm GHz}$. Detailed simulation parameters are summarized in Table~\ref{tab:network_layers}. The dataset used for analysis contains a total of $310,000$ samples, divided into $240,000$ training samples, $60,000$ validation samples, and $10,000$ test samples for model training and performance evaluation. The total number of ports at the FAS side is set to \( N_s = 512 \).

The performance is evaluated through the NMSE, defined as 
\begin{small}\begin{equation}
\text{NMSE}(\hat{\mathbf{g}}, \mathbf{g}) = \frac{\sum_{n=1}^{N_\text{test}} \left\| \mathbf{g}_n - \hat{\mathbf{g}}_n \right\|^2}{\sum_{n=1}^{N_\text{test}} \left\| \mathbf{g}_n \right\|^2},
\end{equation}\end{small}
where \( N_\text{test} \) denotes the number of test samples. The predicted CSI matrix \( \hat{\mathbf{g}} \) is obtained by rearranging the extrapolated tensor \( \hat{\mathbf{U}}_c \) into the standard CSI format. The batch size used in training is set to $32$. The initial learning rate is $0.0005$. For optimization, the AdamW optimizer is employed with the parameters \( \beta_1 = 0.9 \), \( \beta_2 = 0.999 \), and the weight decay set to $0.0001$. We set the maximum strength $\gamma$ to $0.5$, $\mu$ to $0.05$, and the dropout rate to $0.5$, $\beta$ to $0.02$.

\begin{figure}[]
\centering
\subfigure[3.4GHz]{
\includegraphics[width=.75\columnwidth]{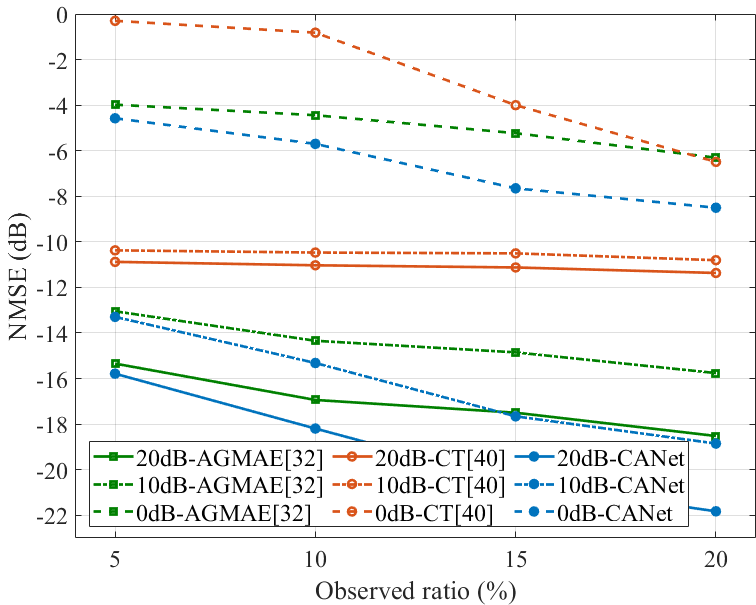}\label{fig:2a_base}
}
\vspace{-2mm}
\subfigure[28GHz]{
\includegraphics[width=.75\columnwidth]{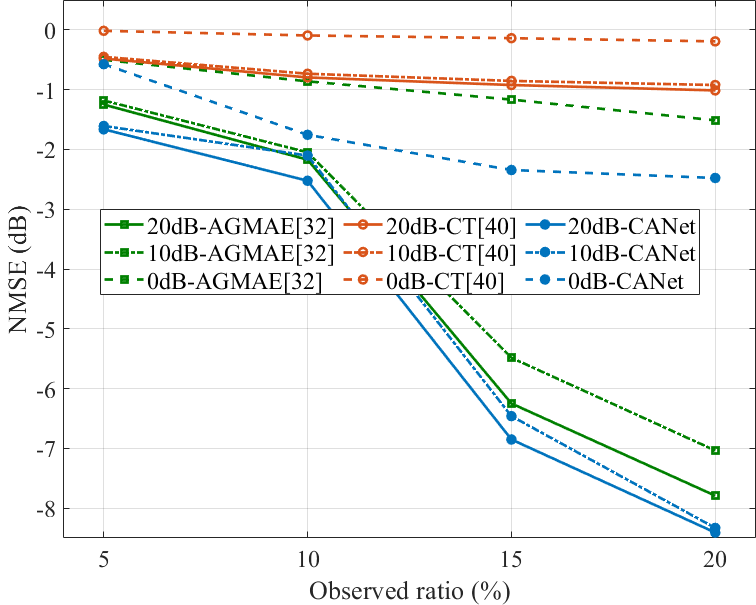}\label{fig:2b_28ghz}
}
\caption{NMSE versus the number of observed ports with different operating frequencies. \textcolor{red}{CANet consistently outperforms benchmark models across all observation ratios, demonstrating superior robustness in reconstructing high-resolution CSI from sparse observations.}} 
\vspace{-2mm}
\end{figure}
\subsection{Performance analysis}
\textcolor{red}{Fig.~2 illustrates the NMSE performance of different models as a function of the observation ratio for a $2~\mathrm{cm}\times4~\mathrm{cm}$ FAS array under multiple SNR conditions, where Figs.~\ref{fig:2a_base} and \ref{fig:2b_28ghz} correspond to the carrier frequencies of $3.4$~GHz and $28$~GHz, respectively. We reproduce AGMAE in \cite{zhang2024learning} as a representative deep-learning-based baseline tailored for FAS scenarios, and additionally include CT\cite{zhang2023ai} as a comparison model, which was originally designed for channel extrapolation in conventional \textbf{multiple-input multiple-output (MIMO)} systems, to evaluate the applicability of generic extrapolation models to FAS settings. In the figure, the same line style is used to indicate the same SNR condition.}

\textcolor{red}{For both carrier frequencies, the NMSE of all models decreases monotonically with the observation ratio, indicating that richer port observations improve CSI reconstruction accuracy. However, under sparse observations, especially in low-SNR conditions, CT exhibits limited performance gains, suggesting that directly reusing MIMO-oriented extrapolation structures is insufficient to capture the fine-grained port correlations inherent in FAS. In contrast, AGMAE demonstrates more stable extrapolation performance across both frequency bands, while the proposed CANet consistently achieves the lowest NMSE under all SNR and observation ratio settings, highlighting its adaptability to different carrier frequencies.}

\begin{figure}[]
\centering

\includegraphics[width=.75\columnwidth]{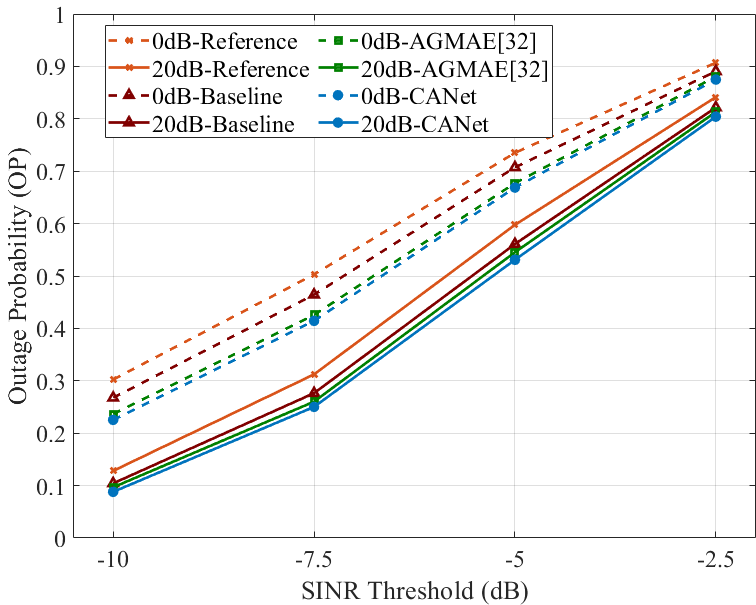}\label{fig:3a_OP}

\caption{Outage probability versus SINR threshold at a fixed observation ratio of $10\%$ under 0 dB and 20 dB observation SNR.  \textcolor{red}{CANet achieves the lowest outage probability among extrapolation schemes, effectively mitigating noise to approach the performance of direct transmission.}} 
\vspace{-2mm}
\end{figure}

\textcolor{red}{To further evaluate the impact of channel extrapolation accuracy on communication performance, the outage probability is evaluated based on the signal model and port selection criteria described in \cite{Waqar2023Deep}. Fig.~3 presents the outage probability (OP) as a function of the SINR threshold under a fixed observation ratio of $10\%$ for different methods. Here, \emph{Baseline} denotes the case where transmission is performed using full but noisy CSI, while \emph{Reference} corresponds to direct transmission based only on the observable port CSI. Figs.~\ref{fig:3a_OP} correspond to observation input SNRs of $0$~dB and $20$~dB, respectively.}

\textcolor{red}{Under both SNR settings, CANet achieves the lowest OP over the entire SINR threshold range. Although the Baseline exploits full-port observations, its performance is still limited by measurement noise; in contrast, CANet effectively suppresses observation noise during extrapolation by learning the spatial correlation and statistical structure of FAS channels, thereby achieving improved communication reliability. As the observation input SNR increases to $20$~dB, the performance gap among different methods narrows, while CANet consistently maintains a clear advantage, demonstrating its robustness under varying noise conditions.}

\begin{table}[t]
\caption{Ablation Study of Physics-Informed Modules 
($2~\mathrm{cm}\times4~\mathrm{cm}$ FAS, 20\% Observation Ratio)}
\label{tab:ablation}
\centering
\setlength{\tabcolsep}{5pt} 
\begin{tabular}{c cc ccc}
\toprule
\textbf{ID} & \multicolumn{2}{c}{\textbf{Modules}} 
            & \multicolumn{3}{c}{\textbf{NMSE (dB)}} \\
\cmidrule(lr){2-3} \cmidrule(lr){4-6}
 & Perturb. & FFT Loss 
 & 0 dB & 10 dB & 20 dB \\
\midrule
1 & -- & -- & $-6.22$ & $-15.88$ & $-18.86$ \\
2 & -- & \checkmark & $-6.51$ & $-16.42$ & $-19.45$ \\
3 & \checkmark & -- & $-7.35$ & $-17.46$ & $-20.47$ \\
\midrule
\textbf{4} & \textbf{\checkmark} & \textbf{\checkmark} 
           & $-\textbf{8.50}$ & $-\textbf{18.85}$ & $-\textbf{21.83}$ \\
\bottomrule
\end{tabular}
\end{table}

\textcolor{red}{Regarding computational efficiency, the proposed CANet consists of 12.40M parameters and requires 2.59 \textbf{giga floating point operations (GFLOPs)} per inference. While slightly more complex than the GAT-based AGMAE (1.27 GFLOPs), CANet significantly reduces the complexity of the standard CT baseline by 49.4$\%$ (from 5.12 to 2.59 GFLOPs). Considering the multi-\textbf{tera operations per second (TOPS)} capability of current 6G-oriented mobile platforms, CANet maintains an excellent balance between extrapolation robustness and deployment feasibility.}
\subsection{Ablation experiments}
\textcolor{red}{To verify the effectiveness of the proposed FAS-motivated components, an ablation study is conducted as shown in Table II. Compared to the basic model (ID 1), the independent inclusion of the FFT loss (ID 2) and the spatial amplitude perturbation strategy (ID 3) both lead to significant NMSE reductions across various SNR levels. Specifically, the full CANet (ID 4) achieves the best performance, with an additional gain of approximately 3 dB at 20 dB SNR compared to ID 1. These results confirm that explicitly enforcing frequency-domain consistency and simulating spatial uncertainty effectively guide the network to capture the intrinsic propagation physics rather than overfitting to pixel-wise noise.}
\section{Conclusions}
In this correspondence, we proposed CANet, a robust CSI extrapolation model that excels in processing unobserved port positions in FAS. CANet integrates a CAB, CSCA, and a ConvNeXt v2 backbone. This combination allows for adaptive CSI feature extraction at both local and global levels. A spatial amplitude perturbation strategy was also proposed to enhance the loss function with an FFT loss. Our simulation results indicated that CANet consistently outperforms benchmark models, demonstrating improved accuracy and remarkable robustness in channel extrapolation under various noise conditions. 

\nocite{*}

\bibliographystyle{IEEEtran}
\bibliography{bibfile}

@article{gao2025enabling,
  title={Enabling {6G} through multi-domain channel extrapolation: {Opportunities} and challenges of generative artificial intelligence},
  author={Gao, Yuan and Lu, Zichen and Wu, Yifan and Jin, Yanliang and Zhang, Shunqing and Chu, Xiaoli and Xu, Shugong and Wang, Cheng-Xiang},
  journal={IEEE Communications Magazine},
  year={2025},
  note      = {early access, \url{doi: 10.1109/MCOM.001.2500246},},
  month=oct}

@article{du2024secure,
  title={Secure task offloading in blockchain-enabled mec networks with improved pbft consensus},
  author={Du, Jianbo and others},
  journal={IEEE Transactions on Cognitive Communications and Networking},
  year={2024},
  publisher={IEEE}
}

@article{hu2025computation,
  title={Computation offloading and resource allocation in mixed cloud/vehicular-fog computing systems},
  author={Hu, Bintao and Du, Jianbo and Zhang, Jie and Chu, Xiaoli},
  journal={IEEE Transactions on Mobile Computing},
  year={2025},
  publisher={IEEE}
}

@article{jiang2025towards,
  title={Towards channel foundation models (CFMs): Motivations, methodologies and opportunities},
  author={Jiang, Jun and Gao, Yuan and Wu, Xinyi and Xu, Shugong},
  journal={arXiv preprint arXiv:2507.13637},
  year={2025}
}

@article{gao2024performance,
  title={On the performance of an integrated communication and localization system: an analytical framework},
  author={Gao, Yuan and others},
  journal={IEEE Transactions on Vehicular Technology},
  volume={73},
  number={7},
  pages={10845--10849},
  year={2024},
  publisher={IEEE}
}

@article{gao2025stochastic,
  title={A Stochastic Geometry-Based Analytical Framework for Integrated Localization and Communication Systems},
  author={Gao, Yuan and others},
  journal={IEEE Internet of Things Journal},
  year={2025},
  publisher={IEEE}
}

@ARTICLE{SSnet2025gao,
author={Gao, Yuan and others},
  journal={IEEE Journal on Selected Areas in Communications}, 
  title={{SSNet}: Flexible and robust channel extrapolation for fluid antenna systems enabled by an self-supervised learning framework}, 
  year={2025},
  volume={},
  number={},
  pages={1-1},
  note      = {early access, \url{doi: 10.1109/JSAC.2025.3619472},},
  month=oct}

@article{xu2025enhanced,
  title={Enhanced Fingerprint-based Positioning With Practical Imperfections: {Deep} learning-based approaches},
  journal={{IEEE} Wirel. Commun.},
  author={Xu, Shugong and others},
    year={2025},
  volume={},
  number={},
  pages={1-1},
  note      = {early access, \url{doi: 10.1109/MWC.2025.3600205},},
  month=oct

}

@article{Wong2022Bruce,
  author    = {K.-K. Wong and K.-F. Tong and Y. Shen and Y. Chen and Y. Zhang},
  title     = {{Bruce Lee}-inspired fluid antenna system: Six research topics and the potentials for {6G}},
  journal   = {Frontiers Commun. Netw.},
  volume    = {3},
  year      = {2022},
  month     = mar,
  note      = {Art. no. 853416}
}

@ARTICLE{OTFSGao2025,
  title={Joint Channel Estimation and Data Detection for {OTFS} Systems: {A} Lightweight Deep Learning Framework With a Novel Data Augmentation Method},
  author={Gao, Yuan and others},
  journal={IEEE Internet of Things Journal},
  year={2025},
  publisher={IEEE}
}

@article{gao2026csiextra,
  title={{AI}-driven channel state information ({CSI}) extrapolation for {6G}: {Current} situations, challenges and future research},
  author={Gao, Yuan and others},
  journal={arXiv preprint arXiv:2601.00159},
  year={2026}
}

@article{New2024Tutorial,
  author    = {W. K. New and others},
  title     = {A tutorial on fluid antenna system for 6G networks: Encompassing communication theory, optimization methods and hardware designs},
  journal   = {IEEE Commun. Surv. Tuts.},
  year      = {2024},
  note      = {early access, \url{doi: 10.1109/COMST.2024.3498855},},
}

@article{I27_basbug2017design,
  author    = {S. Basbug},
  title     = {Design and synthesis of antenna array with movable elements along semicircular paths},
  journal   = {IEEE Antennas Wireless Propag. Lett.},
  volume    = {16},
  pages     = {3059--3062},
  year      = {2017},
  month     = oct
}

@misc{I24_shen2024design,
  author    = {Y. Shen and others},
  title     = {Design and implementation of mmWave surface wave enabled fluid antennas and experimental results for fluid antenna multiple access},
  howpublished = {\url{arXiv:2405.09663}},
  note      = {arXiv preprint},
  year      = {2024},
  month     = may
}

@article{I26_zhang2024pixel,
  author    = {J. Zhang and others},
  title     = {A novel pixel-based reconfigurable antenna applied in fluid antenna systems with high switching speed},
  journal   = {IEEE Open J. Antennas \& Propag.},
  volume    = {6},
  number    = {1},
  pages     = {212--228},
  year      = {2025},
  month     = feb
}

@misc{Liu-2025arxiv,
  author        = {B. Liu and K. F. Tong and K. K. Wong and C.-B. Chae and H. Wong},
  title         = {Be water, my antennas: Riding on radio wave fluctuation in nature for spatial multiplexing using programmable meta-fluid antenna},
  howpublished  = {\url{arXiv:2502.04693}},
  note          = {arXiv preprint},
  year          = {2025},
  month         = feb
}

@article{Lu-2025,
  author    = {W.-J. Lu and others},
  title     = {Fluid antennas: Reshaping intrinsic properties for flexible radiation characteristics in intelligent wireless networks},
  journal   = {IEEE Commun. Mag.},
  volume    = {63},
  number    = {5},
  pages     = {40--45},
  year      = {2025},
  month     = may
}

@article{I22_wong2020perflim,
  author    = {K. K. Wong and A. Shojaeifard and K. F. Tong and Y. Zhang},
  title     = {Performance limits of fluid antenna systems},
  journal   = {IEEE Commun. Lett.},
  volume    = {24},
  number    = {11},
  pages     = {2469--2472},
  year      = {2020},
  month     = nov
}

@article{I20_wong2021FAS,
  author    = {K. K. Wong and A. Shojaeifard and K. F. Tong and Y. Zhang\vspace{0mm}},
  title     = {Fluid antenna systems},
  journal   = {IEEE Trans. Wireless Commun.},
  volume    = {20},
  number    = {3},
  pages     = {1950--1962},
  year      = {2021},
  month     = mar
}

@article{G5_new2023SISO-FAS,
  author    = {W. K. New and K. K. Wong and H. Xu and K. F. Tong and C. B. Chae},
  title     = {Fluid antenna system: New insights on outage probability and diversity gain},
  journal   = {IEEE Trans. Wireless Commun.},
  volume    = {23},
  number    = {1},
  pages     = {128--140},
  year      = {2024},
  month     = jan
}

@article{H7_Espinosa2024Anew,
  author    = {P. Ram\'{i}rez-Espinosa and D. Morales-Jimenez and K. K. Wong},
  title     = {A new spatial block-correlation model for fluid antenna systems},
  journal   = {IEEE Trans. Wireless Commun.},
  volume    = {23},
  number    = {11},
  pages     = {15829--15843},
  year      = {2024},
  month     = nov
}

@article{G7_farshard2024SISO-FAS-Secure,
  author    = {F. Rostami Ghadi and others},
  title     = {Physical layer security over fluid antenna systems: Secrecy performance analysis},
  journal   = {IEEE Trans. Wireless Commun.},
  volume    = {23},
  number    = {12},
  pages     = {18201--18213},
  year      = {2024},
  month     = dec
}

@article{G16_Lai2024FAS-RIS,
  author    = {X. Lai and others},
  title     = {{FAS-RIS}: A block-correlation model analysis},
  journal   = {IEEE Trans. Veh. Technol.},
  volume    = {74},
  number    = {2},
  pages     = {3412--3417},
  year      = {2025},
  month     = feb
}

@article{G27_Zhu2024FA-IM,
  author    = {J. Zhu\vspace{0mm} and others},
  title     = {Index modulation for fluid antenna-assisted MIMO communications: System design and performance analysis},
  journal   = {IEEE Trans. Wireless Commun.},
  volume    = {23},
  number    = {8},
  pages     = {9701--9713},
  year      = {2024},
  month     = aug
}

@article{G26_Zhu2024FAIM-RIS,
  author    = {J. Zhu and others},
  title     = {Fluid Antenna Empowered Index Modulation for {RIS}-Aided {mmWave} Transmissions},
  journal   = {IEEE Trans. Wireless Commun.},
  volume    = {24},
  number    = {2},
  pages     = {1635--1647},
  year      = {2025},
  month     = feb
}

@article{H11_hong2025Downlink,
  author    = {H. Hong and others},
  title     = {Downlink {OFDM-FAMA} in {5G-NR} Systems},
  journal   = {IEEE Trans. Wireless Commun.},
  note      = {\url{arXiv:2501.06974}},
  year      = {2025}
}

@article{Wang2024Fluid,
  author    = {C. Wang and others},
  title     = {Fluid Antenna System Liberating Multiuser {MIMO} for {ISAC} via Deep Reinforcement Learning},
  journal   = {IEEE Trans. Wireless Commun.},
  volume    = {23},
  number    = {9},
  pages     = {10879--10894},
  year      = {2024},
  month     = sep
}

@article{G11_ye2024MIMO-FAS,
  author    = {Y. Ye and others},
  title     = {Fluid Antenna-Assisted {MIMO} Transmission Exploiting Statistical {CSI}},
  journal   = {IEEE Commun. Lett.},
  volume    = {28},
  number    = {1},
  pages     = {223--227},
  year      = {2024},
  month     = jan
}

@article{G13_Efrem2024MIMO-FAS,
  author    = {C. N. Efrem and I. Krikidis},
  title     = {Transmit and Receive Antenna Port Selection for Channel Capacity Maximization in Fluid-{MIMO} Systems},
  journal   = {IEEE Wireless Commun. Lett.},
  volume    = {13},
  number    = {11},
  pages     = {3202--3206},
  year      = {2024},
  month     = nov
}

@article{New2024Information,
  author    = {W. K. New and K.-K. Wong and H. Xu and K.-F. Tong and C.-B. Chae},
  title     = {An Information-Theoretic Characterization of {MIMO-FAS}: Optimization, Diversity–Multiplexing Tradeoff and $q$-Outage Capacity},
  journal   = {IEEE Trans. Wireless Commun.},
  volume    = {23},
  number    = {6},
  pages     = {5541--5556},
  year      = {2024},
  month     = jun
}

@inproceedings{gao2024c2s,
  title={{C2S}: {An} {Transformer}-Based Framework to Extrapolate Sensing Channel From Communication Channel},
  author={Gao, Yuan and Wu, Xinyi and Gao, Yiling and Xu, Shugong},
  booktitle={2024 IEEE 100th Vehicular Technology Conference (VTC2024-Fall)},
  pages={1--5},
  year={2024},
  organization={IEEE}
}

@article{jin2025linformer,
  title={Linformer: A linear-based lightweight transformer architecture for time-aware mimo channel prediction},
  author={Jin, Yanliang and others},
  journal={IEEE Transactions on Wireless Communications},
  year={2025},
  publisher={IEEE}
}

@inproceedings{Li2024Model,
  author      = {G. Li and H. Zhang and C. Wang and B. Wang},
  title       = {Model-Driven Channel Extrapolation for Massive Fluid Antenna},
  booktitle   = {Proc.\ IEEE Int. Conf. Commun. Workshops (ICC Workshops)},
  pages       = {1146--1151},
  year        = {2024},
  month       = jun,
  address     = {Denver, CO, USA},
}

@inproceedings{Wu2024Channel,
  author      = {X. Wu and H. Zhang and C.-C. Wang and Z. Li},
  title       = {Channel State Information Extrapolation in Fluid Antenna Systems Based on Masked Language Model},
  booktitle   = {Proc.\ IEEE Int. Conf. Commun. Workshops (ICC Workshops)},
  pages       = {1383--1388},
  year        = {2024},
  month       = jun,
  address     = {Denver, CO, USA},
}

@article{zhang2024learning,
  author    = {H. Zhang and others},
  title     = {Learning-Induced Channel Extrapolation for Fluid Antenna Systems Using Asymmetric Graph Masked Autoencoder},
  journal   = {IEEE Wireless Commun. Lett.},
  volume    = {13},
  number    = {6},
  pages     = {1665--1669},
  year      = {2024},
  month     = jun
}

@article{Xu2024Channel,
  author    = {H. Xu and others},
  title     = {Channel Estimation for {FAS}-Assisted Multiuser {mmWave} Systems},
  journal   = {IEEE Commun. Lett.},
  volume    = {28},
  number    = {3},
  pages     = {632--636},
  year      = {2024},
  month     = mar
}

@article{New-2025twc,
  author    = {W. K. New and others},
  title     = {Channel Estimation and Reconstruction in Fluid Antenna System: Oversampling Is Essential},
  journal   = {IEEE Trans. Wireless Commun.},
  volume    = {24},
  number    = {1},
  pages     = {309--322},
  year      = {2025},
  month     = jan
}

@article{Xu-2025wcl,
  author    = {B. Xu and Y. Chen and Q. Cui and X. Tao and K. K. Wong},
  title     = {Sparse Bayesian Learning-Based Channel Estimation for Fluid Antenna Systems},
  journal   = {IEEE Wireless Commun. Lett.},
  volume    = {14},
  number    = {2},
  pages     = {325--329},
  year      = {2025},
  month     = feb
}

@article{zhang2023successive,
  author    = {Z. Zhang and J. Zhu and L. Dai and R. W. Heath Jr},
  title     = {Successive Bayesian Reconstructor for Channel Estimation in Fluid Antenna Systems},
  journal   = {IEEE Trans. Wireless Commun.},
  volume    = {24},
  number    = {3},
  pages     = {1992--2006},
  year      = {2025},
  month     = mar
}

@article{Waqar2023Deep,
  author    = {N. Waqar and K.-K. Wong and K.-F. Tong and A. Sharples and Y. Zhang},
  title     = {Deep Learning Enabled Slow Fluid Antenna Multiple Access},
  journal   = {IEEE Commun. Lett.},
  volume    = {27},
  number    = {3},
  pages     = {861--865},
  year      = {2023},
  month     = mar
}

@inproceedings{Woo2023Convnext,
  author      = {S. Woo and others},
  title       = {ConvNeXt {V2}: Co-Designing and Scaling ConvNets With Masked Autoencoders},
  booktitle   = {Proc.\ IEEE/CVF Conf.\ Comput. Vision \& Pattern Recognition (CVPR)},
  pages       = {16133--16142},
  year        = {2023},
  month       = jun,
  address     = {Vancouver, BC, Canada},
}

@inproceedings{Liu2024Universal,
  author      = {C. Liu and Y. Cao and X. Su and H. Zhu},
  title       = {Universal Frequency Domain Perturbation for Single-Source Domain Generalization},
  booktitle   = {Proc.\ ACM Int. Conf. Multimedia},
  pages       = {6250--6259},
  year        = {2024},
  month       = oct,
  address     = {Melbourne, Australia},
}

@inproceedings{Yang2025Single,
  author      = {Z. Yang and C. Yu},
  title       = {A Single Source Generalization Model via Spatial Amplitude Perturbation and Sensitivity Guidance for Colored Medical Image Segmentation},
  booktitle   = {Proc.\ Int. Conf. Pattern Recognition},
  year        = {2024},
  month       = dec,
  address     = {Kolkata, India},
}

@inproceedings{Alr2019Deep,
  author={Alrabeiah, Muhammad and Alkhateeb, Ahmed},
  title={Deep Learning for TDD and FDD Massive MIMO: Mapping Channels in Space and Frequency}, 
  booktitle={2019 53rd Asilomar Conference on Signals, Systems, and Computers}, 
  year={2019},
  pages={1465-1470},
  doi={10.1109/IEEECONF44664.2019.9048929}
}

@ARTICLE{zhang2023ai,
  author={Zhang, Zhen and others},
  journal={IEEE Vehicular Technology Magazine}, 
  title={{AI}-Based Time-, Frequency-, and Space-Domain Channel Extrapolation for {6G}: {Opportunities} and Challenges}, 
  year={2023},
  volume={18},
  number={1},
  pages={29-39},
  doi={10.1109/MVT.2023.3234169}
}

@ARTICLE{Xu2022sparse,
  author={Xu, Xiaowen and others},
  journal={IEEE Transactions on Communications}, 
  title={Sparse Bayesian Learning Based Channel Extrapolation for RIS Assisted MIMO-OFDM}, 
  year={2022},
  volume={70},
  number={8},
  pages={5498-5513},
  doi={10.1109/TCOMM.2022.3184640}
}

@ARTICLE{Zhang2021Deep,
  author={Zhang, Shun and others},
  journal={IEEE Wireless Communications}, 
  title={Deep Learning Based Channel Extrapolation for Large-Scale Antenna Systems: Opportunities, Challenges and Solutions}, 
  year={2021},
  volume={28},
  number={6},
  pages={160-167},
  doi={10.1109/MWC.001.2000534}}
\end{document}